\documentclass[aps,twocolumn,epsfig,prb]{revtex4}
\newcommand{\bsigma}{\mbox{\boldmath $\sigma$}}

\usepackage{amsmath}
\usepackage{graphicx}

\def\nn{\nonumber}

\begin{document}

\title{Chirality dependent frequency shift of radial breathing mode in
metallic carbon nanotubes}


\author{Ken-ichi~Sasaki$^{a}$, Riichiro~Saito$^{a}$, Gene~Dresselhaus$^{b}$,
Mildred~S.~Dresselhaus$^{c,d}$, 
Hootan~Farhat$^{e}$, and Jing~Kong$^{d}$
}

\affiliation{$^{a}$Department of Physics, Tohoku University and CREST, JST, 
Sendai, 980-8578, Japan}
\affiliation{$^{b}$Francis Bitter Magnet Laboratory, 
$^{c}$Department of Physics, 
$^{d}$Department of Electrical Engineering and Computer Science, 
$^{e}$Department of Materials Science and Engineering,
Massachusetts Institute of Technology, Cambridge, MA 02139-4307}

\date{\today}
 
\begin{abstract}
 A phonon frequency shift of the radial breathing mode 
 for metallic single wall carbon nanotubes 
 is predicted as a function of Fermi energy.
 Armchair nanotubes do not show any frequency shift while
 zigzag nanotubes exhibit phonon softening, 
 but this softening is not associated with the broadening.
 This chirality dependence originates from  
 a curvature-induced energy gap and 
 a special electron-phonon coupling mechanism for radial breathing modes.
 Because of the particle-hole symmetry, 
 only the off-site deformation potential contributes to the
 frequency shift. 
 On the other hand, 
 the on-site potential contributes to the Raman intensity,
 and the radial breathing mode intensity is stronger than 
 that of the $G$ band.
 The relationship between the chirality dependence of the frequency
 shift of the radial breathing mode and the $\Gamma$ point optical phonon
 frequency shift is discussed.
\end{abstract}

\maketitle

\section{introduction}

Raman spectroscopy has been widely used 
for the characterization of 
carbon nanotube~\cite{r642,w699,yu01,l818,doorn04,i1049} 
and graphene~\cite{ferrari06,yan07,pimenta07} samples, 
since Raman spectroscopy is a non-destructive 
and non-contact measurement that can be carried out 
under ambient air pressure and at room temperature.
The $G$ band which gives a strong Raman intensity 
at around $1600{\rm cm}^{-1}$ ($\simeq$0.2eV) in graphene 
consists of the $\Gamma$ point 
longitudinal optical (LO) and transverse optical (TO) phonon
modes.~\cite{saito98book} 
For single wall carbon nanotubes (SWNTs),
these two phonon modes with $A_{1g}$ symmetry split 
by the curvature effect into two Raman features
which we call the $G^+$ and $G^-$ bands.~\cite{jorio02}
It has been observed for metallic SWNTs
that the peak positions of the $G^+$ and $G^-$ features
shift differently from each other 
as a function of the Fermi energy,
$E_{\rm F}$.~\cite{farhat07,nguyen07,wu07,das07}
It is known that 
virtual electron-hole pair creation by the electron-phonon (el-ph)
interaction gives a self-energy correction to the phonon frequency, 
which is relevant to the phonon frequency shift.
The behavior of the $G^+$ and $G^-$ bands indicates that 
the el-ph interactions for the $\Gamma$ point LO/TO modes appear 
to be different from each other in metallic SWNTs. 
The frequency shift for the LO/TO mode for graphene
and nanotubes has been discussed by
many
authors.~\cite{dubay02,piscanec04,lazzeri06prl,ishikawa06,popov06,lazzeri06prb,caudal07,calandra07,piscanec07,ando08}
In a previous paper,
we showed that the phonon softening for the LO and TO phonon 
is chirality dependent in which the curvature effect is
important.~\cite{q1239} 
For example, only armchair SWNTs do not exhibit any frequency
shift of the TO mode due to the absence of a curvature-induced energy
gap in armchair SWNTs.

A similar effect may be observed for 
the radial breathing mode (RBM)
which appears in the SWNT Raman spectra 
at around $240{\rm cm}^{-1}$ ($\simeq$30meV)
for a nanotube diameter of around $d_t \simeq 1$nm.
The RBM is often used to assign the diameter and chirality of a
nanotube.~\cite{l818,z1066,telg04}  
For metallic SWNTs except for armchair SWNTs,
the curvature of a cylindrical graphene layer induces 
a small energy gap.~\cite{saito92prb,ouyang01,guelseren02}
Since the curvature-induced energy gap 
has a similar energy to that for the RBM,
especially for SWNTs with diameter less than $2$nm,
the curvature-induced energy gap may affect the
phonon softening of the RBM, 
which is the motivation of the present paper.
Thus, it is important to estimate this frequency shift of 
the RBM for metallic SWNTs.
In this paper, 
we examine the dependence of the RBM frequency shift
on $E_{\rm F}$ and chirality for metallic SWNTs.
The relationship between the RBM and the LO/TO modes
is examined for the el-ph matrix element
of the electron-hole pair creation.

Since the frequency shift of a phonon mode 
is mainly due to a low energy electron-hole pair creation,
a theory for electrons near the Fermi energy 
is useful for calculating 
the matrix element for an electron-hole pair creation.
The el-ph interaction 
has previously been examined in terms of an effective-mass model
by several authors.~\cite{suzuura02,ishikawa06}
It is known that
the el-ph interaction consists of on-site and off-site terms.
The on-site term represents a process whereby 
a $\pi$-electron changes its energy due to a 
deformation potential but stays at the same position.
The off-site term on the other hand represents a scattering process
whereby a $\pi$-electron moves into a nearest carbon site due to
a deformation potential.~\cite{jiang05prb}
As for the $\Gamma$ point LO/TO modes,
the el-ph interaction is given only by the off-site term.
In a previous paper, we showed that
the off-site term of 
the el-ph interaction becomes sensitive to 
the position of the cutting line 
or to the curvature effect.~\cite{q1239}
In this paper, we will show that 
an interesting property of the RBM is that 
the el-ph interaction is given not only by the off-site term
but also by the on-site term.
Although the on-site term does not contribute 
to the frequency shift of the RBM 
because of the particle-hole symmetry about the Dirac point,
the on-site term of the el-ph interaction
enhances the Raman intensity of the RBM.

This paper is organized as follows.
In Sec.~\ref{sec:rbm_cig} we show that
the curvature-induced energy gap determines 
the basic features for the frequency shift of the RBM.
In Sec.~\ref{sec:hami}, 
we show the el-ph interaction for the RBM 
using an effective-mass model 
for a graphene sheet with a lattice deformation.
In Sec.~\ref{sec:main},
we calculate the frequency shift of the RBM 
as a function of the Fermi energy.
We compare our results with experimental data
in Sec.~\ref{sec:exp}.
A discussion and summary 
of these effects are given in Sec.~\ref{sec:dis}.
The effective-mass Hamiltonian used in Sec.~\ref{sec:hami}
and Sec.~\ref{sec:main} is derived in Appendix~\ref{app:W+A}
and Appendix~\ref{app:chi}, respectively.

\section{radial breathing mode and the curvature-induced energy
 gap}\label{sec:rbm_cig}

\begin{figure}[htbp]
 \begin{center}
  \includegraphics[scale=0.5]{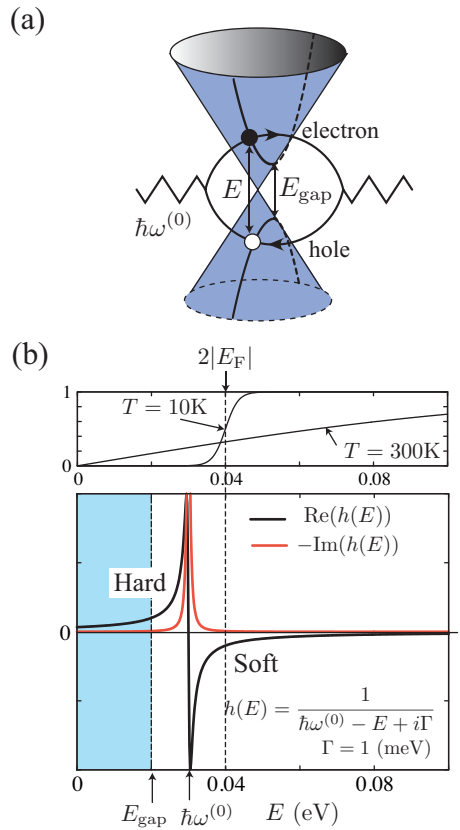}
 \end{center}
 \caption{(Color online) (a) 
 An intermediate electron-hole pair
 consists of an electron in the conduction band
 and a hole in the valence band.
 The RBM is denoted by a zigzag line and an electron-hole
 pair is represented by a loop.
 A cutting line for a metallic SWNT is shown as the solid curve 
 projected on the surface of the Dirac cone.
 The electron-hole pair creation is possible 
 only when $E \ge E_{\rm gap}$.
 (b) The energy correction to the phonon energy 
 by an intermediate electron-hole pair,
 especially the sign of ${\rm Re}(h(E))$ (black curve),
 that is, frequency hardening or softening,
 depends on the energy of the intermediate state $E$.
 The contribution to $\hbar \omega^{(2)}$
 of a low energy electron-hole pair 
 satisfying $0 \le E \le E_{\rm gap}$ 
  is forbidden.
 ${\rm Im}(h(E))$ (red curve)
 is nonzero only when $E$ is very close to $\hbar \omega^{(0)}$,
 which shows that the RBM can resonantly 
 decay into an electron-hole pair 
 with the same energy.
 The figure is the case of $\hbar \omega^{(0)}=30$meV, 
 $E_{\rm gap}=20$meV.
 The Fermi distribution function 
 $f[E_{\rm h}({\bf k})-E_{\rm F}]- f[E_{\rm e}({\bf k})-E_{\rm F}]$
 is plotted for $T=10$K and $T=300$K as a function of $E$
 in the case of $|E_{\rm F}|=20$meV.
 }
 \label{fig:2ndpert}
\end{figure}

Here we show that
the basic features of the frequency shift of the RBM
are determined by the curvature-induced energy gap.
A renormalized RBM energy 
becomes $\hbar \omega= \hbar \omega^{(0)}+\hbar \omega^{(2)}$
where $\omega^{(0)}$ is the unperturbed RBM frequency
and $\omega^{(2)}$ is the quantum correction 
to the RBM frequency due to the el-ph interaction.
We assume that
$\omega^{(0)}$ is a monotonic function of the tube diameter ($d_t$[nm])
and is modeled as linear in inverse diameter with an offset as 
\begin{align}
 \omega^{(0)} = \frac{c_1}{d_t} + c_2,
 \label{eq:rbm_ene}
\end{align}
where $c_1=223.5$[cm$^{-1}$] and $c_2=12.5$[cm$^{-1}$]
are experimentally derived parameters as obtained 
by Bachilo {\it et al.},~\cite{strano03,bachilo02} while
$\hbar \omega^{(2)}$ is calculated by
second-order perturbation theory~\cite{q1239} as
\begin{align}
 \hbar \omega^{(2)} = 
 & 2 \sum_{\bf k}
 \frac{|\langle {\rm eh}({\bf k})|
 {\cal H}_{\rm ep}|\omega^{(0)}
 \rangle|^2}{\hbar\omega^{(0)}-[E_{\rm e}({\bf k})-E_{\rm h}({\bf
 k})]+i\Gamma}  \nn \\
 & \times 
 \left\{
 f[E_{\rm h}({\bf k})-E_{\rm F}]-
 f[E_{\rm e}({\bf k})-E_{\rm F}] \right\}.
 \label{eq:omega_2}
\end{align}
In Eq.~(\ref{eq:omega_2}),
the factor 2 comes from spin degeneracy,
$E_{\rm e}({\bf k})$ ($E_{\rm h}({\bf k})$)
is the energy of an electron (hole) with wave vector ${\bf k}$,
$\langle {\rm eh}({\bf k})| {\cal H}_{\rm
ep}|\omega^{(0)} \rangle$ is the el-ph matrix element that 
the RBM changes into an electron-hole pair 
with wave vector ${\bf k}$
(see Fig.~\ref{fig:2ndpert}(a)), and
$f[x]$ is the Fermi function.
We obtain $\Gamma$ (the half of the decay width) 
by calculating $\Gamma= - {\rm Im} (\hbar \omega^{(2)})$
self-consistently in Eq.~(\ref{eq:omega_2}).

We consider the real part and imaginary part of 
the denominator of Eq.~(\ref{eq:omega_2}),
$h(E)=1/(\hbar \omega^{(0)}-E+i\Gamma)$, as a function of 
the energy of an intermediate electron-hole pair state,
$E \equiv E_{\rm e}({\bf k})-E_{\rm h}({\bf k})$
(see Fig.~\ref{fig:2ndpert}(b)).
${\rm Re}(h(E))$ has a positive (negative) value
when $E < \hbar \omega^{(0)}$ ($E > \hbar \omega^{(0)}$), and
the lower (higher) energy electron-hole pair
makes a positive (negative) contribution to 
$\hbar \omega^{(2)}$.
Therefore, 
the sign of the contribution to ${\rm Re} (\hbar \omega^{(2)})$,
i.e., frequency hardening or softening, depends on its electron-hole
virtual state energy, $E$.
The curvature-induced energy gap, $E_{\rm gap}$, 
affects the RBM frequency shift
since an electron-hole pair creation is possible 
only when $E \ge E_{\rm gap}$.
When $0 < E_{\rm gap} \le \hbar \omega^{(0)}$,
the contribution to frequency hardening in Eq.~(\ref{eq:omega_2})
is suppressed.
When $E_{\rm gap} > \hbar \omega^{(0)}$,
not only are all the positive contributions to the RBM frequency 
suppressed, but some negative contributions are also suppressed.
Further,
${\rm Im}(h(E))$ is nonzero only when $E$ is very close to 
$\hbar \omega^{(0)}$,
which shows that the RBM phonon can resonantly 
decay into an electron-hole pair 
with the same energy.
This means that 
$\Gamma \simeq 0$ when $E_{\rm gap} > \hbar \omega^{(0)}$ 
because no electron-hole pair excitation is allowed
near $E=\hbar \omega^{(0)}$.
Thus, it is important to compare 
the values of $E_{\rm gap}$ and $\hbar \omega^{(0)}$
for each $(n,m)$ SWNT.

\begin{figure}[htbp]
 \begin{center}
  \includegraphics[scale=0.6]{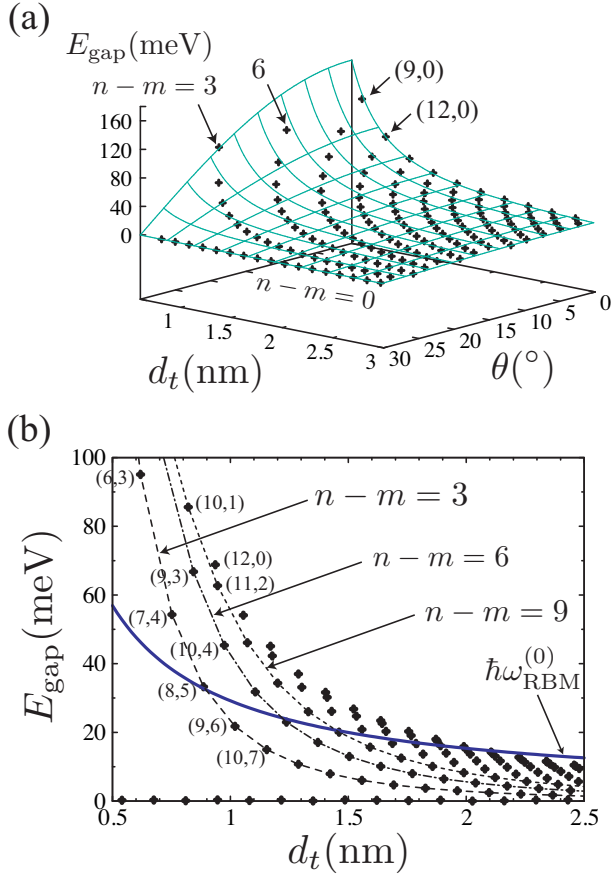}
 \end{center}
 \caption{(Color online)
 (a)
 The dependence of the curvature-induced energy gap, $E_{\rm gap}$,
 on the chiral angle $\theta$ and tube diameter $d_t$.
 The surface is a plot of Eq.~(\ref{eq:Egap}) 
 which reproduces well the calculated results.
 (b) The $d_t$ dependence of $E_{\rm gap}$, 
 which is given as a one-dimensional projection of (a)
 onto the $d_t$ axis.
 We plot the energy of the RBM, $\hbar \omega^{(0)}$ 
 of Eq.~(\ref{eq:rbm_ene}), as a solid blue curve for comparison.
 The points on the dashed, dot-dashed, and dotted curves 
 satisfy $n-m=3,6,9$, respectively.
 }
 \label{fig:Erbm}
\end{figure}

In Fig.~\ref{fig:Erbm}(a) we plot $E_{\rm gap}$
for each $(n,m)$ for metallic SWNTs
as a function of the chiral angle $\theta(^\circ)$ and tube diameter $d_t$(nm).
We performed the energy band structure calculation 
in an extended tight-binding framework~\cite{samsonidze04}
to obtain $E_{\rm gap}$.
Figure~\ref{fig:Erbm}(a) shows that, for a fixed $d_t$,
a zigzag SWNT ($\theta=0^\circ$)
has the largest value of $E_{\rm gap}$
and an armchair SWNT ($\theta=30^\circ$) has no energy gap.
The $(n,m)$ values associated with a curve are given by
$n-m=3q$ ($q=0,1,2,\cdots$).
The calculated results are well reproduced by
\begin{align}
 E_{\rm gap} = \frac{c}{d_t^2} \cos 3\theta,
 \label{eq:Egap}
\end{align}
with $c=60$(meV$\cdot$nm$^2$).~\cite{kane97}
Using Eqs.~(\ref{eq:rbm_ene}) and (\ref{eq:Egap}) 
for zigzag SWNTs ($\theta=0^\circ$),
we see that $E_{\rm gap}$ is larger than $\hbar \omega^{(0)}$
when $d_t \alt 2$nm (see Fig.~\ref{fig:Erbm}(b)).
Thus, no decay process (or spectral broadening)
contributing to $\Gamma$ is available
for a zigzag SWNT with $d_t \alt 2$nm.
On the other hand,
a decay process contributing to $\Gamma$ is available
regardless of the chirality of a metallic SWNT with $d_t \agt 2$nm.
For example,
in Fig.~\ref{fig:Erbm}(b),
we see that a $(12,0)$ tube 
does not exhibit a finite value of $\Gamma$
even though we get a finite value for the el-ph matrix element,
but $(9,6)$ and $(10,7)$ may exhibit a broadening.

In addition, 
because of the Fermi distribution function $f[x]$ in
Eq.~(\ref{eq:omega_2}),
an electron-hole pair satisfying $E < 2|E_{\rm F}|$ 
can not contribute to the energy shift
at zero temperature due to the Pauli principle.
At zero temperature, 
when $|E_{\rm F}| \simeq \hbar \omega^{(0)}/2$,
then $\hbar \omega^{(2)}$ takes a minimum value 
since all positive contributions to $\hbar \omega^{(2)}$
are suppressed in Eq.~(\ref{eq:omega_2}).~\cite{q1239}
$E_{\rm F}$ and $E_{\rm gap}$ play a very similar role 
at zero temperature, but their difference becomes clear 
at a finite temperature. 
For example, when 
$E_{\rm gap} < \hbar \omega^{(0)}$,
then $\Gamma$ can be nonzero even when 
$2|E_{\rm F}| > \hbar \omega^{(0)}$.
On the other hand, 
when $E_{\rm gap} > \hbar \omega^{(0)}$,
then $\hbar \omega^{(2)}$ does not have an imaginary part ($\Gamma = 0$)
even at room temperature, 
regardless of the value of $E_{\rm F}$.
This difference between $E_{\rm F}$ and $E_{\rm gap}$
is understood in Eq.~(\ref{eq:omega_2}) as
$\langle {\rm eh}({\bf k})| {\cal H}_{\rm ep}|\omega^{(0)}
\rangle=0$ for $E \le E_{\rm gap}$ and
as that
$f[E_{\rm h}({\bf k})-E_{\rm F}]- f[E_{\rm e}({\bf k})-E_{\rm
F}]$ is not zero 
for $E < 2 |E_{\rm F}|$ (at a finite temperature)
as shown in Fig.~\ref{fig:2ndpert}(b).
Thus, $E_{\rm gap}$ determines 
whether a SWNT can exhibit a broadening.

\section{electron-phonon interaction by effective-mass
 theory}\label{sec:hami} 

Next we show the el-ph interaction in the Hamiltonian 
by effective-mass theory,
which is used for calculating the matrix element 
of the el-ph interaction.
We will derive Eqs.~(\ref{eq:H0K}), (\ref{eq:W}) and (\ref{eq:A}) 
from the nearest-neighbor tight-binding Hamiltonian 
in Appendix~\ref{app:W+A}.~\cite{sasaki05}

\subsection{Unperturbed Hamiltonian}

The unperturbed Hamiltonian in the effective-mass model
for $\pi$-electrons near the K point of a graphene sheet 
is given by
\begin{align}
 {\cal H}^{\rm K}_0 = v_{\rm F} \bsigma \cdot \hat{\mathbf{p}},
 \label{eq:H0K}
\end{align}
where $v_{\rm F}$ is the Fermi velocity,
$\hat{\bf p}=-i\hbar \nabla$ is the momentum operator,
and $\bsigma=(\sigma_x,\sigma_y)$ is the Pauli matrix.
${\cal H}^{\rm K}_0$ is a $2\times 2$ matrix 
which operates on the two component wavefunction:
\begin{align}
 \psi^{\rm K}({\bf r}) = 
 \begin{pmatrix}
  \psi^{\rm K}_{\rm A}({\bf r})\cr
  \psi^{\rm K}_{\rm B}({\bf r})
 \end{pmatrix}
\end{align}
where $\psi^{\rm K}_{\rm A}({\bf r})$ and $\psi^{\rm K}_{\rm B}({\bf r})$
are the wavefunctions of $\pi$-electrons
for the sublattices A and B, respectively, around the K point.
The energy eigenvalue of Eq.~(\ref{eq:H0K})
is given by $\pm v_{\rm F}|{\bf p}|$ and 
the energy dispersion relation
shows a linear dependence at the Fermi
point,~\cite{saito98book} which is known as the Dirac cone.
The eigenstates for $E=+v_{\rm F}|{\bf p}|$ and 
$E=-v_{\rm F}|{\bf p}|$ are given by
\begin{align}
 \begin{split}
  & \psi^{\rm K}_{c,{\bf k}}({\bf r})
  = \frac{e^{i{\bf k}\cdot {\bf r}}}{\sqrt{2S}}
  \begin{pmatrix}
   e^{-i\Theta({\bf k})/2} \cr
   e^{+i\Theta({\bf k})/2}
  \end{pmatrix}, \\
  & \psi^{\rm K}_{v,{\bf k}}({\bf r})
  = \frac{e^{i{\bf k}\cdot {\bf r}}}{\sqrt{2S}}
  \begin{pmatrix}
   e^{-i\Theta({\bf k})/2} \cr
   -e^{+i\Theta({\bf k})/2}
  \end{pmatrix},
 \end{split}
 \label{eq:wfK}
\end{align}
which are a conduction state and a valence state 
with wavevector ${\bf k}$, respectively.
In Eq.~(\ref{eq:wfK}),
$S$ denotes the surface area of graphene, 
the wavevector is measured from the K point, and
$\Theta({\bf k})$ is defined by an angle of 
${\bf k}=(k_x,k_y)$ measured from the $k_x$ axis as
$(k_x,k_y) \equiv |{\bf k}| 
(\cos \Theta({\bf k}), \sin \Theta({\bf k}))$.
Here the $k_y$ axis is defined by the direction 
of a zigzag nanotube axis
(see the coordinate system in Fig.~\ref{fig:graphene}(a)).
The energy eigenstate for the valence band,
$\psi^{\rm K}_{v,{\bf k}}({\bf r})$ is given by 
$\sigma_z \psi^{\rm K}_{c,{\bf k}}({\bf r})$.
This results from the particle-hole symmetry of the Hamiltonian:
$\sigma_z {\cal H}^{\rm K}_0 \sigma_z 
=-{\cal H}_0^{\rm K}$.
The dynamics of $\pi$-electrons near the K' point relates to 
the electrons near the K point by time-reversal symmetry,
$\psi^{\rm K} \to (\psi^{\rm K'})^*$.
Because lattice vibrations do not break 
time-reversal symmetry,
we only consider the electrons near the K point in this paper.

\subsection{Perturbation}

\begin{figure}[htbp]
 \begin{center}
  \includegraphics[scale=0.5]{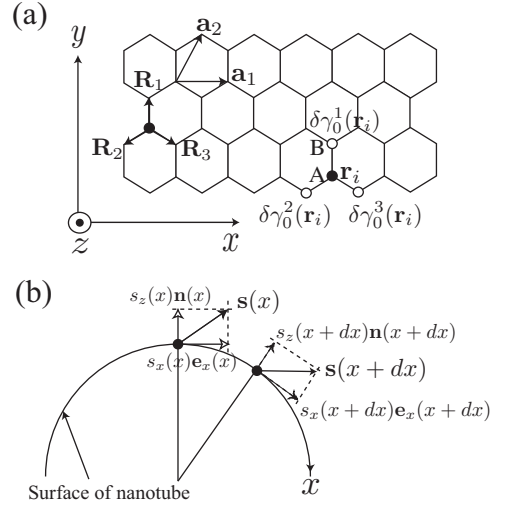}
 \end{center}
 \caption{(a)
 A hexagonal unit cell of graphene consists of 
 {\rm A} (closed circle) and {\rm B} (open circle) atoms.
 ${\bf a}_1$ and ${\bf a}_2$ are lattice vectors.
 ${\bf R}_a$ ($a=1,2,3$)
 are vectors pointing to the nearest-neighbor  
 {\rm B} sites from an {\rm A} site.
 For the coordinate system of $(x,y)$, 
 the ${\bf R}_a$ ($a=1,2,3$) are written, respectively, as
 ${\bf R}_1=a_{\rm cc}{\bf e}_y$,
 ${\bf R}_2=-(\sqrt{3}/2)a_{\rm cc}{\bf e}_x -(1/2)a_{\rm cc}{\bf e}_y$,
 and 
 ${\bf R}_3=(\sqrt{3}/2)a_{\rm cc}{\bf e}_x -(1/2)a_{\rm cc}{\bf e}_y$
 where $a_{\rm cc}$ is the carbon-carbon bond length and
 ${\bf e}_x$ (${\bf e}_y$)
 is the dimensionless unit vector for the $x$-axis ($y$-axis).
 Local modulations of the hopping integral are defined by 
 $\delta \gamma^a_0({\bf r}_i)$ ($a=1,2,3$) where ${\bf r}_i$ 
 is the position of an A-atom.
 (b)
 The displacement vector for the RBM, ${\bf s}(x)$, 
 is decomposed in terms of the normal $s_z(x)$ and 
 tangential $s_x(x)$ components.
 The derivative of the normal unit vector ${\bf n}(x)$ with respect to
 $x$ gives a component along ${\bf e}_x(x)$, which modifies the net
 displacement along the $x$ direction.
 }
 \label{fig:graphene}
\end{figure}

Lattice deformation modifies
the nearest-neighbor hopping integral locally 
as $-\gamma_0 \to -\gamma_0+\delta \gamma_0^a({\bf r}_i)$ 
($a=1,2,3$) (see Fig.~\ref{fig:graphene}(a)).
The corresponding perturbation of the lattice deformation 
is given by
\begin{align}
 {\cal H}_1 \equiv 
 \sum_{i \in {\rm A}} \sum_{a=1,2,3} 
 \delta \gamma^a_0(\mathbf{r}_i) 
 \left[
 (c_{i+a}^{\rm B})^\dagger c_i^{\rm A} + 
 (c_i^{\rm A})^\dagger c_{i+a}^{\rm B}
 \right],
 \label{eq:H1}
\end{align}
where $c_i^{\rm A}$ is the annihilation operator of 
a $\pi$ electron of an A-atom at position ${\bf r}_i$,
and $(c^{\rm B}_{i+a})^\dagger$ is a creation operator
of position ${\bf r}_{i+a}$ $(={\bf r}_i+{\bf R}_a)$
of a B-atom where ${\bf R}_a$ ($a=1,2,3$) are vectors 
pointing to the three nearest-neighbor B sites from an A site
(see Fig.~\ref{fig:graphene}(a)).

The perturbation of Eq.~(\ref{eq:H1})
gives rise to scattering within a 
region near the K point (intravalley
scattering) whose interaction
is given by a deformation-induced gauge field 
${\bf A}({\bf r})=(A_x({\bf r}),A_y({\bf r}))$
in Eq.~(\ref{eq:H0K}) as 
\begin{align}
 v_{\rm F} \bsigma \cdot \left[ \hat{\mathbf{p}}+{\bf A}({\bf r}) \right].
 \label{eq:W}
\end{align}
${\bf A}({\bf r})$ is defined from 
$\delta \gamma^a_0({\bf r})$ ($a=1,2,3$) as~\cite{sasaki05,sasaki06jpsj}
\begin{align}
 \begin{split}
  & v_{\rm F} A_x({\bf r}) = \delta \gamma^1_0({\bf r})
  - \frac{1}{2} \left[ \delta \gamma^2_0({\bf r}) +
  \delta \gamma^3_0({\bf r}) \right], \\
  & v_{\rm F} A_y({\bf r}) = \frac{\sqrt{3}}{2} 
  \left[ \delta \gamma^2_0({\bf r}) -
  \delta \gamma^3_0({\bf r}) \right].
 \end{split}
 \label{eq:A}
\end{align}
When $\delta \gamma^2_0=\delta \gamma^3_0=0$,
then ${\bf A}({\bf r})=(A_x({\bf r}),0)$ and 
${\bf A}({\bf r})\cdot {\bf R}_1=0$.
Similarly, when $\delta \gamma^1_0=\delta \gamma^3_0=0$,
we have ${\bf A}({\bf r})\cdot {\bf R}_2=0$.
Generally, the direction of ${\bf A}({\bf r})$
is pointing perpendicular to the bond 
whose hopping integral is changed from $\gamma_0$.

When the displacement vector of a carbon atom at ${\bf r}$ is
${\bf s}({\bf r})=(s_x({\bf r}),s_y({\bf r}),s_z({\bf r}))$, 
the perturbation to the nearest-neighbor hopping integral 
is given by
\begin{align}
 \delta \gamma^a_0({\bf r}) 
 = \frac{g_{\rm off}}{\ell a_{\rm cc}}
 {\bf R}_a \cdot
 \{ {\bf s}({\bf r}+{\bf R}_a)- {\bf s}({\bf r}) \},
 \label{eq:delta_gamma}
\end{align}
where $g_{\rm off}$ is the off-site coupling constant and 
$\ell=3a_{\rm cc}/2$.
By expanding ${\bf s}({\bf r}+{\bf R}_a)$ in a Taylor series around
${\bf s}({\bf r})$ as
${\bf s}({\bf r}+{\bf R}_a)={\bf s}({\bf r})+({\bf R}_a\cdot
\nabla) {\bf s}({\bf r}) + \cdots$, we approximate
Eq.~(\ref{eq:delta_gamma}) as
\begin{align}
 \delta \gamma^a_0({\bf r}) 
 \simeq \frac{g_{\rm off}}{\ell a_{\rm cc}}
 {\bf R}_a \cdot 
 \left\{ ({\bf R}_a \cdot \nabla) {\bf s}({\bf r}) \right\}.
 \label{eq:dg_ac}
\end{align}
Putting ${\bf R}_1=a_{\rm cc}{\bf e}_y$,
${\bf R}_2=-(\sqrt{3}/2)a_{\rm cc}{\bf e}_x -(1/2)a_{\rm cc}{\bf e}_y$,
and ${\bf R}_3=(\sqrt{3}/2)a_{\rm cc}{\bf e}_x -(1/2)a_{\rm cc}{\bf
e}_y$, into the right-hand side of Eq.~(\ref{eq:dg_ac}), 
we obtain the corresponding deformation-induced gauge field
of Eq.~(\ref{eq:A}) as
\begin{align}
 \begin{split}
  & v_{\rm F}A_x({\bf r}) = \frac{g_{\rm off}}{2} \left[
  -\frac{\partial s_x({\bf r})}{\partial x}+
  \frac{\partial s_y({\bf r})}{\partial y} \right], \\
  & v_{\rm F}A_y({\bf r}) = \frac{g_{\rm off}}{2} \left[
  \frac{\partial s_x({\bf r})}{\partial y}+ 
  \frac{\partial s_y({\bf r})}{\partial x} \right].
 \end{split}
 \label{eq:A_ac}
\end{align}
Further
the displacements of carbon atoms give 
an on-site deformation potential
\begin{align}
 {\cal H}_{\rm on} = 
 g_{\rm on} \sigma_0 
 \left[
 \frac{\partial s_x({\bf r})}{\partial x} + 
 \frac{\partial s_y({\bf r})}{\partial y}
 \right].
 \label{eq:On_ac}
\end{align}
Here $\sigma_0$ is the $2\times 2$ unit matrix and 
$\partial s_x({\bf r})/\partial x + \partial s_y({\bf r})/\partial y$
represents the change of the area of a graphene sheet.~\cite{suzuura02} 
In Eqs.~(\ref{eq:A_ac}) and (\ref{eq:On_ac}),
according to the density functional calculation 
by Porezag {\it et al.},~\cite{porezag95}
we adopt the off-site coupling constant
$g_{\rm off}=6.4$eV and the on-site coupling constant
$g_{\rm on} =17.0$eV.~\cite{jiang05prb,sasaki07local}

Since Eqs.~(\ref{eq:A_ac}) and (\ref{eq:On_ac})
are proportional to the derivatives of 
$s_x({\bf r})$ and $s_y({\bf r})$,
that is, they are proportional to ${\bf q}$,
the el-ph matrix element for the in-plane longitudinal/transverse 
acoustic (LA/TA) phonon modes vanishes 
at the $\Gamma$ point
where ${\bf q}$ is the phonon wave vector.
Namely,
${\bf A}({\bf r})=0$ 
and ${\cal H}_{\rm on}=0$
in the limit of ${\bf q}=0$.
Among the TA phonon modes,
there is an out-of-plane TA (oTA) phonon mode.
The oTA mode shifts carbon atoms on the flat 2D graphene sheet 
into the $z$-direction (see Fig.~\ref{fig:graphene}(a) and (b)).
The oTA mode of graphene corresponds to the RBM of a nanotube
even though the RBM is not an acoustic phonon mode.~\cite{saito98book}
In the following, we will show that 
the el-ph interaction for the RBM is enhanced 
due to the curvature of the nanotube
as compared with the oTA mode of graphene
since the RBM is a bond-stretching mode 
because of the cylindrical structure
of SWNTs.

The displacements of the RBM modify 
the radius of a nanotube as 
$r \to r + s_z({\bf r})$ (see Fig.~\ref{fig:graphene}(b)).
A change of the radius gives rise to two effects 
on the electronic state.
One effect is a shift of the wavevector around the tube axis.
The distance between two wavevectors around the tube axis depends on 
the inverse of the radius due to the periodic boundary condition
and a change of the radius results in a shift of the wavevector.
The other effect is that 
the RBM can change the area on the surface of the nanotube
even at the $\Gamma$ point.
This results in an enhancement of the on-site interaction.
These two effects are relevant to the fact that the normal vector 
on the surface of a nanotube is pointing in a different direction 
depending on the position.
To show this, we take a (zigzag) nanotube as shown in 
Fig.~\ref{fig:graphene}(b).
Let us denote the displacement vectors 
of two carbon atoms at $x$ and $x+dx$
as ${\bf s}(x)$ and ${\bf s}(x+dx)$, then
an effective length for the displacement along the $x$ axis
between the nearest two atoms is given by
\begin{align}
 D_x={\bf e}_x(x+dx) \cdot \left[
 {\bf s}(x+dx)- {\bf s}(x) \right].
 \label{eq:dis}
\end{align}
By decomposing ${\bf s}(x)$
in terms of a normal and a tangential unit vector as
${\bf s}(x)=s_z(x){\bf n}(x)+s_x(x){\bf e}_x(x)$
(see Fig.~\ref{fig:graphene}(b)),
we see that Eq.~(\ref{eq:dis}) becomes
\begin{align}
 D_x 
 &= s_x(x+dx) + s_z(x+dx) {\bf e}_x(x+dx) \cdot {\bf n}(x+dx) \nn \\
 &- s_x(x) {\bf e}_x(x+dx) \cdot {\bf e}_x(x) - 
 s_z(x){\bf e}_x(x+dx) \cdot {\bf n}(x) \nn \\
 & = dx \left\{
 \frac{\partial s_x(x)}{\partial x}+ \frac{s_z(x)}{r} \right\} + \cdots,
 \label{eq:rbm_sz}
\end{align}
where we have used the following equations:
\begin{align}
 \begin{split}
  & {\bf n}(x+dx) 
  = {\bf n}(x) + \frac{dx}{r} {\bf e}_x(x) + \cdots, \\
  & {\bf e}_x(x+dx) 
  = {\bf e}_x(x) + \frac{dx}{r} {\bf n}(x) + \cdots.
 \end{split}
\end{align}
Equation~(\ref{eq:rbm_sz}) shows that
the net displacement along the $x$ axis is modified
by the curvature of the nanotube as 
$\partial_x s_x({\bf r}) \to \partial_x s_x({\bf r})+s_z({\bf r})/r$.
The correction is negligible
for a graphene sheet ($r \to \infty$).

The el-ph interaction for the RBM is included by replacing 
$\partial_x s_x({\bf r})$ with $\partial_x s_x({\bf r})+s_z({\bf r})/r$
in Eqs.~(\ref{eq:A_ac}) and (\ref{eq:On_ac}).
In Eq.~(\ref{eq:A_ac}), we have an additional 
deformation-induced gauge field,
$v_{\rm F}A_x({\bf r})=-(g_{\rm off}/2)(s_z({\bf r})/r)$,
for the RBM mode which gives rise to a shift of the wavevector around the
tube axis.
In Eq.~(\ref{eq:On_ac}), 
it is shown that the RBM produces on-site deformation potential 
of $g_{\rm on} \sigma_0 (s_z({\bf r})/r)$.
Finally, 
we obtain the el-ph interaction 
for the $\Gamma$ point (${\bf q}=0$: ${\bf s}({\bf r})$ is a constant) RBM, 
as
\begin{align}
 {\cal H}_{\rm ep} 
 &= -\frac{g_{\rm off}}{2} \frac{s_z}{r}\sigma_x  + 
 g_{\rm on} \frac{s_z}{r} \sigma_0 \nn \\
 &= \frac{2s_z}{d_t}
 \begin{pmatrix}
  g_{\rm on} & -\frac{g_{\rm off}}{2} \cr
  -\frac{g_{\rm off}}{2} & g_{\rm on}
 \end{pmatrix}.
 \label{eq:Hep}
\end{align}

\section{Electron-Phonon Matrix element and frequency
 shift}\label{sec:main}

In this section
we calculate the el-ph matrix element 
for an electron-hole pair creation
and the corresponding frequency shift of the RBM.

\begin{figure}[htbp]
 \begin{center}
  \includegraphics[scale=0.5]{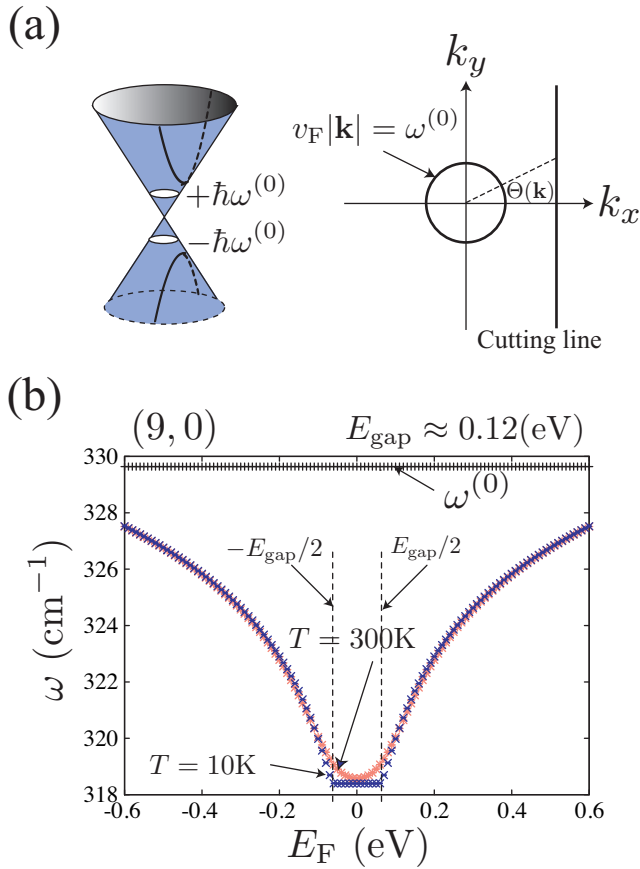}
 \end{center}
 \caption{(Color online) 
 (a)
 The decay width is zero because of 
 $E_{\rm gap} > \hbar \omega^{(0)}$.
 This is shown by the relative positions of the cutting line 
 and the equi-energy contour circle satisfying 
 $v_{\rm F}|{\bf k}|=\omega^{(0)}$ with respect to 
 the Dirac point.
 (b) 
 The $E_{\rm F}$ dependence of the RBM frequency 
 in the case of the $(9,0)$ zigzag SWNT 
 at room temperature (red curve) 
 and at 10 K (blue curve).
 We plot $\omega^{(0)}$ as the black line for comparison.
 }
 \label{fig:zig_EF}
\end{figure}

From Eqs.~(\ref{eq:wfK}) and (\ref{eq:Hep}),
the el-ph matrix element for an electron-hole pair generation 
near the K point is given by
\begin{align}
 &\langle {\rm eh}({\bf k})|{\cal H}_{\rm ep}
 |\omega^{(0)} \rangle \nn \\
 &= \int 
 (\psi^{\rm K}_{c,{\bf k}}({\bf r}))^\dagger
 {\cal H}_{\rm ep}
 \psi^{\rm K}_{v,{\bf k}}({\bf r}) d^2{\bf r}
 \nonumber \\
  &= \frac{s_z}{d_t}
 \begin{pmatrix}
  e^{+i\frac{\Theta({\bf k})}{2}} \cr
  e^{-i\frac{\Theta({\bf k})}{2}}
 \end{pmatrix}^t
 \begin{pmatrix}
  g_{\rm on} & -\frac{g_{\rm off}}{2} \cr
  -\frac{g_{\rm off}}{2} & g_{\rm on}
 \end{pmatrix} 
 \begin{pmatrix}
  e^{-i\frac{\Theta({\bf k})}{2}} \cr 
  -e^{+i\frac{\Theta({\bf k})}{2}}
 \end{pmatrix} \nn \\
 &=ig_{\rm off}\frac{s_z}{d_t}\sin \Theta({\bf k}).
 \label{eq:elph_RBM}
\end{align}
It is noted that
the $g_{\rm on}$ term does not contribute to 
the RBM frequency shift.
This is because
the particle-hole symmetry:
$\psi_{v,{\bf k}}({\bf r}) = \sigma_z \psi_{c,{\bf k}}({\bf r})$,
gives a vanishing matrix element,
$\psi^\dagger_{c,{\bf k}}({\bf r}) g_{\rm on}\sigma_0 
\psi_{v,{\bf k}}({\bf r})= g_{\rm on} 
\psi^\dagger_{c,{\bf k}}({\bf r}) \sigma_0 \sigma_z 
\psi_{c,{\bf k}}({\bf r})=0$
in Eq.~(\ref{eq:elph_RBM}).
We note that
$\sin \Theta({\bf k})$ of Eq.~(\ref{eq:elph_RBM}) 
indicates that
low energy electron-hole pairs near the Dirac point 
(${\bf k}$-states satisfying $\Theta({\bf k}) \approx 0$
on the cutting line, 
see Fig.~\ref{fig:zig_EF}(a))
are hardly excited.
Instead, high energy electron-hole pairs 
(${\bf k}$-states of $\Theta({\bf k}) \approx \pm \pi/2$)
do contribute to the frequency softening.
Putting Eq.~(\ref{eq:elph_RBM}) into Eq.~(\ref{eq:omega_2}), 
we calculate the frequency shift 
as a function of $E_{\rm F}$ for a (9,0) zigzag SWNT.
In Fig.~\ref{fig:zig_EF}(b), 
we plot $\omega^{(0)}$ (black line) and $\omega$ 
for room temperature (red curve) and 
$\omega$ for 10 K (blue curve). 
The frequency difference between 
$E_{\rm F}=0.6$ eV and the Dirac point ($E_{\rm F}=0$)
is about 10 cm$^{-1}$.

It is useful to compare Eq.~(\ref{eq:elph_RBM}) with 
the amplitude of electron-hole pair creation 
by the $\Gamma$ point LO/TO phonon modes
in order to understand the diameter dependence of 
the RBM frequency shift of a zigzag SWNT.
In a previous paper,~\cite{q1239} 
we obtained for zigzag SWNTs that
\begin{align}
 \begin{split}
  & \langle {\rm eh}({\bf k})|
  {\cal H}_{\rm ep}|\omega_{\rm LO} \rangle
  =-ig_{\rm off} \frac{u_{\rm LO}}{a_{\rm cc}} \sin \Theta({\bf k}), \\
  & \langle {\rm eh}({\bf k})|
  {\cal H}_{\rm ep}|\omega_{\rm TO} \rangle
  =-ig_{\rm off} \frac{u_{\rm TO}}{a_{\rm cc}} \cos \Theta({\bf k}),
 \end{split}
 \label{eq:coupling}
\end{align}
where $u_{\rm LO}$ ($u_{\rm TO}$) 
is the amplitude of the LO (TO) phonon mode.
The $\Theta({\bf k})$ dependence of 
the matrix element of Eq.~(\ref{eq:elph_RBM})
is the same as the $\Gamma$ point LO mode 
of Eq.~(\ref{eq:coupling}).
By comparing Eq.~(\ref{eq:elph_RBM}) with Eq.~(\ref{eq:coupling}),
we find that the ratio of the el-ph matrix element squared is given by
\begin{align}
 R =
 \frac{|\langle {\rm eh}({\bf k})|{\cal H}_{\rm ep} |\omega^{(0)}
 \rangle|^2}{|\langle {\rm eh}({\bf k})|{\cal H}_{\rm ep}|\omega_{\rm
 LO} \rangle|^2} =  
 \left( \frac{s_z}{d_t} 
 \frac{a_{\rm cc}}{u_{\rm LO}}\right)^2.
 \label{eq:raito_oprbm}
\end{align}
Since the phonon amplitude
is proportional to the phonon frequency as
$s_z \propto \sqrt{\hbar/M\omega^{(0)}}$ and
$u_{\rm LO} \propto \sqrt{\hbar/M\omega_{\rm LO}}$ 
($M$ is the mass of the carbon atom),
we obtain from Eq.~(\ref{eq:raito_oprbm}) that
\begin{align}
 R \simeq \frac{a_{\rm cc}}{d_t}.
 \label{eq:Rfac}
\end{align}
Here, we have used Eq.~(\ref{eq:rbm_ene}) and 
$\omega_{\rm LO} = 1600{\rm cm}^{-1}$ to get
$s_z^2/u_{\rm LO}^2=\omega_{\rm LO}/\omega^{(0)}
\simeq 7.2(d_t/1{\rm nm})$.
Because the squared matrix element appears in the numerator 
of Eq.~(\ref{eq:omega_2}),
the RBM frequency shift is proportional to the inverse of $d_t$.
This is consistent with the fact that 
the RBM frequency shift of a $(9,0)$ SWNT
is around 10${\rm cm}^{-1}$ because 
the frequency shift of the $\Gamma$ point LO mode~\cite{q1239} 
reaches around 50${\rm cm}^{-1}$
and $R \simeq 0.2$.
The frequency difference between $E_{\rm F}=0.6$eV and the Dirac point
is about 5 cm$^{-1}$ for a (18,0) zigzag SWNT
($d_t\simeq 1.4$nm), whose $d_t$ is twice the $d_t$ of a (9,0) SWNT.
It is expected that the frequency shift of the RBM 
in zigzag SWNTs is expressed by $50(a_{\rm cc}/d_t)$cm$^{-1}$.

In Fig.~\ref{fig:zig_EF}(b),
the decay width is zero 
because there is no electron-hole pair 
satisfying $E < \hbar \omega^{(0)}$
(see Figs.~\ref{fig:zig_EF}(a) and~\ref{fig:Erbm}(b)).
In principle, the decay width due to the el-ph interaction
is absent for zigzag SWNTs when $d_t \alt 2$nm
because the curvature-induced energy gap is larger than 
the original RBM phonon energy:
$E_{\rm gap}\gg \hbar \omega^{(0)}$.
On the other hand, when $d_t > 2$nm,
we have checked that the $\Gamma$ value
is less than 1cm$^{-1}$ due to the small matrix element for 
a larger diameter zigzag SWNT.

For a general $(n,m)$ SWNT
with a chiral angle $\theta$,
the el-ph interaction for the RBM becomes
\begin{align}
 {\cal H}_{\rm ep}(\theta) =
 \frac{2s_z}{d_t}
 \begin{pmatrix}
  g_{\rm on} & -\frac{g_{\rm off}}{2} e^{+i3\theta} \cr
  -\frac{g_{\rm off}}{2} e^{-i3\theta} & g_{\rm on}
 \end{pmatrix},
 \label{eq:rbm_int}
\end{align}
which is derived in Appendix~\ref{app:chi}.
As a result, 
the matrix element for an electron-hole pair creation 
is chirality dependent as
\begin{align}
 \langle {\rm eh}({\bf k})|{\cal H}_{\rm ep}(\theta)
 |\omega^{(0)} \rangle =
 ig_{\rm off}\frac{s_z}{d_t}\sin (\Theta({\bf k})+3\theta),
 \label{eq:elph_RBM_g}
\end{align}
where $k_1 = |{\bf k}|\cos \Theta({\bf k})$
is the wavevector in the direction around the tube axis
and $k_2 = |{\bf k}|\sin \Theta({\bf k})$
is the one along the tube axis.
Thus the frequency shift of the RBM has a chiral angle dependence.
In particular, armchair SWNTs ($\theta=30^\circ$)
do not exhibit any frequency shift
because the matrix element becomes
\begin{align}
 \langle {\rm eh}({\bf k})|{\cal H}_{\rm ep}(\theta) 
 |\omega^{(0)} \rangle =
 ig_{\rm off}\frac{s_z}{d_t}\cos (\Theta({\bf k})),
 \label{eq:aa}
\end{align}
which is zero for a cutting line of the metallic band:
$\Theta({\bf k})=\pm \pi/2$.
This $\Theta({\bf k})$ dependence of Eq.~(\ref{eq:aa})
is the same as that of the TO phonon mode of Eq.~(\ref{eq:coupling})
and the absence of the frequency shift of 
the RBM in armchair SWNTs
is similar to that of the $\Gamma$ point TO mode 
in armchair SWNTs.~\cite{q1239}

\begin{figure}[htbp]
 \begin{center}
  \includegraphics[scale=0.5]{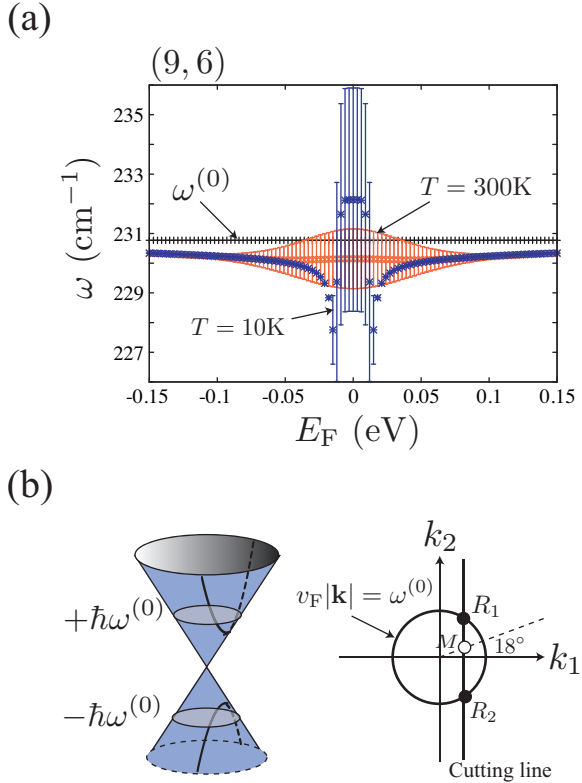}
 \end{center}
 \caption{(Color online)
 (a)
 The $E_{\rm F}$ dependence of the RBM frequency 
 in the case of the $(9,6)$ chiral SWNT 
 at room temperature (red curve) 
 and at 10 K (blue curve).
 We plot $\omega^{(0)}$ as the black line for comparison.
 (b)
 The decay width ($\Gamma$) is non-zero because of 
 $E_{\rm gap} < \hbar \omega^{(0)}$.
 At the $R_1$ and $R_2$ points, there is the contribution to $\Gamma$.
 At $M$ ($\Theta({\bf k})=18^\circ$) the contribution to
 phonon softening becomes a maximum.
 }
 \label{fig:chi_EF}
\end{figure}

In Fig.~\ref{fig:chi_EF}(a)
we show the frequency shift in a $(9,6)$ SWNT
($\theta=24^\circ$).
The factor of $\sin (\Theta({\bf k})+3\theta)$ 
in Eq.~(\ref{eq:elph_RBM_g})
indicates that low energy electron-hole pairs 
satisfying $\Theta({\bf k}) \approx \pi/2 - 3\theta=18^\circ$
on the cutting line
contribute significantly to the frequency shift.
The decay width is non-zero 
because an electron-hole pair can be excited 
near $E \simeq \hbar \omega^{(0)}$
(see Figs.~\ref{fig:chi_EF}(b) and~\ref{fig:Erbm}(b)).

Our numerical calculation shows that
the RBM of a (14,2) chiral SWNT
exhibits an $E_{\rm F}$ dependent frequency shift 
of about 3${\rm cm}^{-1}$ within 
an $E_{\rm F}$ range of $\pm 0.3$eV.
This result is consistent with the experimental result by
Nguyen {\it et al.}~\cite{nguyen07}, who observed
that the RBM of a (14,2) chiral tube 
exhibits a small gate dependent frequency up-shift
up to 3${\rm cm}^{-1}$ within the gate voltage range of $\pm1$eV,
if we assume that the change of the gate voltage of $\pm 1$V
corresponds to $E_{\rm F}\simeq \pm 0.3$eV, namely, 
the gate efficiency factor is about 0.3.

\subsection{Raman Intensity}

We note that 
Eq.~(\ref{eq:Rfac}) does not mean that 
the Raman intensity of the RBM is much smaller than
that of the LO/TO phonon modes.
The Raman intensity is relevant to the el-ph matrix element 
for a photo-excited electron within the conduction states,
\begin{align}
 &\langle {\rm ee}({\bf k})|{\cal H}_{\rm ep}
 |\omega^{(0)} \rangle \nn \\
 &= \int 
 (\psi^{\rm K}_{c,{\bf k}}({\bf r}))^\dagger
 {\cal H}_{\rm ep}
 \psi^{\rm K}_{c,{\bf k}}({\bf r}) d^2{\bf r}
 \nonumber \\
  &= \frac{s_z}{2r}
 \begin{pmatrix}
  e^{+i\frac{\Theta({\bf k})}{2}} \cr
  -e^{-i\frac{\Theta({\bf k})}{2}}
 \end{pmatrix}^t
 \begin{pmatrix}
  g_{\rm on} & -\frac{g_{\rm off}}{2} e^{+i3\theta} \cr
  -\frac{g_{\rm off}}{2} e^{-i3\theta} & g_{\rm on}
 \end{pmatrix} 
 \begin{pmatrix}
  e^{-i\frac{\Theta({\bf k})}{2}} \cr 
  -e^{+i\frac{\Theta({\bf k})}{2}}
 \end{pmatrix} \nn \\
 &=  \frac{s_z}{d_t} 
 \left(
 2g_{\rm on}+g_{\rm off} \cos(\Theta({\bf k})+3\theta)
 \right).
 \label{eq:ee_rbm}
\end{align}
In this case, 
the $g_{\rm on}$ term does contribute to 
the matrix element and enhances the Raman intensity.
Although Eq.~(\ref{eq:ee_rbm}) is chirality dependent,
the dependence is small
since $2g_{\rm on} \gg g_{\rm off}$.
We compare this result with 
the corresponding LO mode el-ph matrix element 
from a conduction state to a conduction state 
in zigzag or armchair SWNTs
near the K point,~\cite{q1239} 
\begin{align}
 \langle {\rm ee}({\bf k})|{\cal H}_{\rm ep}|\omega_{\rm LO} \rangle
 = i g_{\rm off} \frac{u_{\rm LO}}{a_{\rm cc}} 
 \cos \Theta({\bf k}).
 \label{eq:ee_op}
\end{align}
The ratio between Eqs.~(\ref{eq:ee_rbm}) and (\ref{eq:ee_op}) becomes
\begin{align}
 \frac{|\langle {\rm ee}({\bf k})|{\cal H}_{\rm ep}
 |\omega^{(0)} \rangle|}{|\langle {\rm ee}({\bf k})|{\cal H}_{\rm
 ep}|\omega_{\rm LO}\rangle|}
 \approx \frac{g_{\rm on}}{g_{\rm off}} \frac{a_{\rm cc}}{r} 
 \frac{s_z}{u_{\rm LO}}
 \approx 5 \sqrt{\frac{a_{\rm cc}}{d_t}},
\end{align}
which means that 
the intensity of the RBM can be comparable to that of the $G$ band
because the intensity ratio is given by
\begin{align}
 \frac{I_{\rm RBM}}{I_{\rm G}} \equiv 
  \frac{|\langle {\rm ee}({\bf k})|{\cal H}_{\rm ep}
 |\omega^{(0)} \rangle|^2}{|\langle {\rm ee}({\bf k})|{\cal H}_{\rm
 ep}|\omega_{\rm LO}\rangle|^2}
 \simeq \frac{3.5({\rm nm})}{d_t}.
\end{align}

When $\pi$-electrons satisfying $\theta({\bf k}_{ii})\approx 0$
contribute to the intensity most effectively,
the intensity of the RBM is maximum 
for zigzag nanotubes ($\theta =0$)
and is minimum for armchair nanotubes ($\theta=\pi/6$).
The same tendency is observed by the experiment 
of Strano {\it et al.}~\cite{strano03} and 
is confirmed by a first-principles calculation by
Machon {\it et al.}~\cite{machon05}
On the other hand,
the frequency shift of the RBM is absent for armchair nanotubes
while that for zigzag nanotubes is a maximum.
%

\section{experiment}\label{sec:exp}

\begin{figure}[htbp]
 \begin{center}
  \includegraphics[scale=0.8]{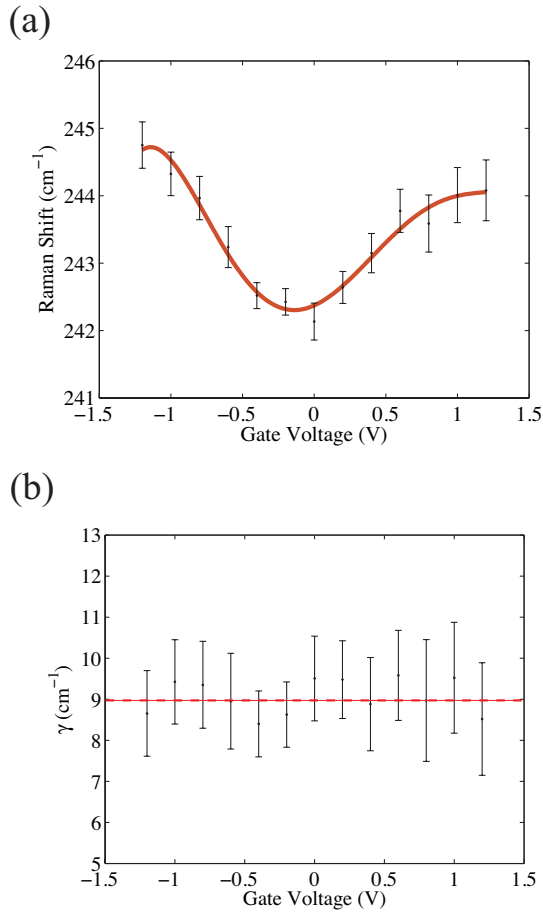}
 \end{center}
 \caption{(Color online)
 (a) The RBM frequency and (b) the spectral width
 as a function of the gate voltage for an isolated SWNT.
 The laser excitation energy is 2.15eV.
 The $(n,m)$ value of the SWNT is assigned as $(11,2)$ or $(12,0)$.
 }
 \label{fig:11_2}
\end{figure}

We next compare our theoretical results 
with experimental Raman RBM data 
taken for an individual SWNT as
a function of the gate voltage 
with the laser excitation energy 
2.15eV.
In Fig.~\ref{fig:11_2}(a), 
we show a plot of the RBM Raman frequency 
as a function of the gate voltage.
It should be noted that the $E_{ii}$ values seem
to be modified slightly as a function of the gate voltage,
which changes the Raman intensity. 
The experimental details 
relevant to Fig.~\ref{fig:11_2}
will be reported elsewhere.~\cite{hootan08}

Here we note that 
the Raman frequency in Fig.~\ref{fig:11_2}(a)
shows a minimum value at zero gate voltage,
while the spectral width in Fig.~\ref{fig:11_2}(b)
appears to be constant within the error bars.
The gate voltage independent offset of the spectra width
does not originate from the 
RBM phonon self-energy due to the electron-hole pair creation
but rather from the life-time of a photo-excited carrier.
The absence of a gate voltage dependent broadening 
shows that $E_{\rm gap}$ is larger than $\hbar \omega^{(0)}$.
This is consistent with the fact that 
the $(n,m)$ for this SWNT 
is assigned as $(11,2)$ or $(12,0)$ using 
a Kataura plot based on the extended tight-binding
model.~\cite{samsonidze04} 
In fact, as shown in Fig.~\ref{fig:Erbm}(b),
the calculated $E_{\rm gap}$ 
(62meV for $(11,2)$, 68meV for $(12,0)$)
is larger than the RBM energy ($\hbar \omega^{(0)} 
\simeq 30$meV).
The RBM frequency difference 
between the gate voltage of 0V and of 1V is
about 2.5 cm$^{-1}$,
which is close to our theoretical estimation for 
a $(11,2)$ or $(12,0)$ SWNT if 
the gate coupling efficiency factor of this SWNT 
is about 0.2.
We estimated the gate coupling efficiency to be about 0.2
for this sample using Eq.~(29) in Ref.~\onlinecite{caudal07}, 
which is an estimation 
based the voltage window in which the $G^-$ broadens.~\cite{hootan08}
In this case, a frequency shift is estimated to be 
about 2.4cm$^{-1}$ for a $(12,0)$ SWNT 
from $(E_{\rm F}/0.6{\rm eV})50(a_{\rm cc}/d_t)$cm$^{-1}$
with $E_{\rm F}=0.2$eV and $d_t\simeq 0.93$nm.
The sample does not exhibit a perfect symmetric shape 
for the RBM frequency 
about positive and negative gate voltage values
but shows some asymmetric shape.
This asymmetry can not be explained by
the phonon self-energy.
We think that the asymmetry is due to 
a change of a spring force constant 
by doping.

\section{discussion and summary}\label{sec:dis}

Let us discuss the similarity between 
the RBM and the LO/TO phonon modes for achiral SWNTs.
The RBM matrix element of Eq.~(\ref{eq:elph_RBM_g}) shows that 
the $\Theta({\bf k})$ dependence
is the same as the LO (TO) matrix element of Eq.~(\ref{eq:coupling})
when $\theta=0^\circ$ ($\theta=30^\circ$).
As a result, 
we obtain using Eq.~(\ref{eq:Rfac}) that 
\begin{align}
 \omega^{(2)}_{\rm RBM} \simeq \omega^{(2)}_{\rm LO} 
 \frac{a_{\rm cc}}{d_t},
\end{align}
for zigzag SWNTs.
This correspondence originates from the character of the RBM 
as an optical phonon mode like the LO/TO modes; that is, 
the A-atom and B-atom oscillate 
in the opposite circumferential directions.
For chiral SWNTs ($\theta \ne 0^\circ$ or $30^\circ$),
on the other hand,
the relationship between the RBM and the LO/TO 
is not as straightforward as that for achiral SWNTs because 
the LO/TO phonon eigenvector is not pointing along 
either the nanotube axis or the circumference.~\cite{reich01prb} 
This changes the $\Theta({\bf k})$ dependence of 
the LO (TO) matrix element of Eq.~(\ref{eq:coupling}) as
\begin{align}
 \begin{split}
  & \langle {\rm eh}({\bf k})|
  {\cal H}_{\rm ep}|\omega_{\rm LO} \rangle
  =-ig_{\rm off} \frac{u_{\rm LO}}{a_{\rm cc}} \sin 
  \left(
  \Theta({\bf k}) + \phi \right), \\
  & \langle {\rm eh}({\bf k})|
  {\cal H}_{\rm ep}|\omega_{\rm TO} \rangle
  =-ig_{\rm off} \frac{u_{\rm TO}}{a_{\rm cc}} \cos 
  \left(
  \Theta({\bf k})+ \phi \right),
 \end{split}
 \label{eq:coupling_g}
\end{align}
where $\phi$ is a parameter for the phonon eigenvector.~\cite{q1239} 
If we consider the frequency shift 
for $|E_{\rm F}| \agt \hbar \omega_{\rm LO(TO)}/2$, 
only high-energy electron-hole pairs contribute to the
frequency shift. 
Then, by rewriting Eq.~(\ref{eq:elph_RBM_g}) using
Eq.~(\ref{eq:coupling_g}), and putting the result into
Eq.~(\ref{eq:omega_2}), 
we obtain using Eq.~(\ref{eq:Rfac}) that 
\begin{align}
 \omega^{(2)}_{\rm RBM} \simeq \omega^{(2)}_{\rm LO} 
 \frac{a_{\rm cc}}{d_t}
 \cos^2 \left( 3\theta - \phi \right),
 \label{eq:LO-RBM}
\end{align}
where we assume $\phi \le 30^\circ$ 
which results in 
$|\omega^{(2)}_{\rm TO}| \ll |\omega^{(2)}_{\rm LO}|$.
We note that
the relationship between the frequency shift of the RBM and that of an
optical phonon mode, which is similar to Eq.~(\ref{eq:LO-RBM}), 
was also derived by Nisoli {\it et al.}~\cite{nisoli07}

We have shown that the electron-hole pair creation is given by the
off-site deformation potential, whose effect is represented by 
the deformation-induced gauge field, ${\bf A}({\bf r})$. 
It is naturally expected that
electron-hole pair creation is enhanced 
where the deformation-induced gauge field appears.
A static ${\bf A}({\bf r})$ field appears 
near the boundary (edge) of the sample.~\cite{sasaki06jpsj}
We will study the effect of the edge on the phonon frequency shift 
in the future.

Finally we discuss the effect of impurities on the RBM frequency.
When impurities are approximated by adding $V({\bf r})\sigma_0$ 
to the effective-mass Hamiltonian of Eq.~(\ref{eq:H0K}),
then it can be shown that the broadening of the RBM due to
the el-ph interaction
is not enhanced by the presence of impurities.
It is because of this that 
the wavefunction of a conduction state relates to 
the wavefunction of a valence state by multiplying $\sigma_z$
(particle-hole symmetry),
and therefore the matrix element of $V({\bf r})\sigma_0$
between a conduction state and a valence state vanishes.
In contrast, the Coulomb interaction 
between an electron and a hole may contribute to 
the broadening of the RBM since an electron and a hole
in the intermediate state
are attracted to each other, 
which affects the life time of the RBM.



In summary, 
for a fixed diameter metallic tube, 
a zigzag SWNT exhibits the maximum RBM frequency shift
and an armchair SWNT does not show any frequency shift.
This is due to the chirality dependent el-ph interaction 
for the RBM.
For a zigzag SWNT, the frequency softening 
is about 10 cm$^{-1}$ in a $(9,0)$ SWNT and it is proportional to the
inverse of the diameter.
When $d_t \alt 2$nm, 
no broadening of the RBM spectra 
appears, since the curvature-induced energy gap
is larger than the RBM phonon energy.
$(9,6)$ and $(10,7)$ SWNTs are candidates which can exhibit a broadening
since the curvature-induced energy gap is smaller
than the RBM energy.
Although the frequency shift of the RBM 
is much smaller than that of the $\Gamma$ point optical phonon mode,
the frequency shift of the RBM 
as a function of the Fermi energy shows a characteristic
behavior depending on the relative position of the cutting line with
respect to the Dirac point.

\section*{Acknowledgment}

R. S. acknowledges a Grant-in-Aid (Nos. 16076201 and 20241023) from
MEXT. MIT authors acknowledge support under NSF Grant DMR 07-04197.

\appendix 

\section{Effective mass theory with lattice deformation}\label{app:W+A}

In this Appendix
we derive Eqs.~(\ref{eq:W}) and (\ref{eq:A}) 
from the nearest-neighbor tight-binding Hamiltonian 
of a graphene sheet with a lattice deformation.

When a graphene sheet does not have any lattice deformation,
the Hamiltonian of $\pi$-electrons is modeled by
\begin{align}
 {\cal H}_0 = 
 -\gamma_0 \sum_{i \in {\rm A}} \sum_{a=1,2,3} 
 \left(
 (c_{i+a}^{\rm B})^\dagger c_i^{\rm A} + 
 (c_i^{\rm A})^\dagger c_{i+a}^{\rm B}
 \right).
 \label{eq:H0}
\end{align}
We use the Bloch theorem to diagonalize Eq.~(\ref{eq:H0}).
The Bloch wavefunction with wavevector ${\bf k}$ is defined by
\begin{align}
 |\Psi_s^{\bf k} \rangle = 
 \frac{1}{\sqrt{N_u}} \sum_{i \in s} 
 e^{i {\bf k} \cdot {\bf r}_i} (c^s_i)^\dagger |0 \rangle \ \ \
 (s = {\rm A},{\rm B}),
 \label{eq:Bloch}
\end{align}
where $N_u$ is the number of hexagonal unit cells
and $|0 \rangle$ denotes the state of carbon atoms 
without $\pi$-electrons.
The off-site matrix element of ${\cal H}_0$ is given by
\begin{align}
 \begin{split}
  & \langle \Psi_{\rm A}^{\bf k} |{\cal H}_0| 
  \Psi_{\rm B}^{\bf k} \rangle
  = - \gamma_0 \sum_{a=1,2,3} f_a({\bf k})
  = - \gamma_0 f({\bf k}), \\
  & \langle \Psi_{\rm B}^{\bf k} |{\cal H}_0|
  \Psi_{\rm A}^{\bf k} \rangle
  = - \gamma_0 \sum_{a=1,2,3} f_a({\bf k})^*
  = - \gamma_0 f({\bf k})^*,
 \end{split}
 \label{eq:H0_bb}
\end{align}
where $f_a({\bf k})\equiv e^{i{\bf k} \cdot {\bf R}_a}$
and $f({\bf k})\equiv \sum_{a=1,2,3} f_a({\bf k})$,~\cite{saito98book}
and the on-site matrix element of ${\cal H}_0$,
$\langle \Psi_{s}^{\bf k}|{\cal H}_0|\Psi_{s}^{\bf k}\rangle$
($s=$A,B), can be taken as zero.
The energy eigenequation is written 
in $2\times 2$ matrix form as
\begin{align}
 E({\bf k})
 \begin{pmatrix}
  | \Psi_{\rm A}^{\bf k} \rangle \cr
  | \Psi_{\rm B}^{\bf k} \rangle
 \end{pmatrix}
 = -\gamma_0 
 \begin{pmatrix}
  0 & f({\bf k}) \cr
  f({\bf k})^* & 0 
 \end{pmatrix}
 \begin{pmatrix}
  | \Psi_{\rm A}^{\bf k} \rangle \cr
  | \Psi_{\rm B}^{\bf k} \rangle
 \end{pmatrix}.
 \label{eq:H_AB}
\end{align}
The energy band structure of ${\cal H}_0$ is obtained 
by solving 
\begin{align}
 \det
 \begin{pmatrix}
  E({\bf k}) & \gamma_0 f({\bf k}) \cr
  \gamma_0 f({\bf k})^* & E({\bf k})
 \end{pmatrix}
 =0.
\end{align}
The solution, $E({\bf k})=+\gamma_0 |f({\bf k})|$ 
($-\gamma_0 |f({\bf k})|$),
is the conduction (valence) energy band.
The conduction energy band and the valence energy band 
touch each other where $|f({\bf k})|$ vanishes.
$|f({\bf k})|=0$ is satisfied at the K point,
${\bf k}_{\rm F}$ ($=(4\pi/3\sqrt{3}a_{\rm cc},0)$), and 
at the K' point, $-{\bf k}_{\rm F}$.
The K or K' point is referred to as the Dirac point.


By expanding $f_a({\bf k})$ in Eq.~(\ref{eq:H0_bb})
around the wavevector of ${\bf k}_{\rm F}$ (the K point),
we obtain $f_a({\bf k}_{\rm F}+{\bf k})=f_a({\bf k}_{\rm F})
+if_a({\bf k}_{\rm F}){\bf k}\cdot {\bf R}_a+\cdots$.
Using ${\bf k}_{\rm F}=(4\pi/3\sqrt{3}a_{\rm cc},0)$,
we get $f_1({\bf k}_{\rm F}) = 1$, 
$f_2({\bf k}_{\rm F})=e^{-i\frac{2\pi}{3}}$, and
$f_3({\bf k}_{\rm F})=e^{+i\frac{2\pi}{3}}$.
Substituting these into Eq.~(\ref{eq:H0_bb}),
we obtain
\begin{align}
 \begin{split}
  & \langle \Psi_{\rm A}^{{\bf k}_{\rm F}+{\bf k}} | 
  {\cal H}_0|\Psi_{\rm B}^{{\bf k}_{\rm F}+{\bf k}} \rangle = 
  \gamma_0 \frac{3a_{\rm cc}}{2} (k_x - ik_y)
  + \cdots, \\
  & \langle \Psi_{\rm B}^{{\bf k}_{\rm F}+{\bf k}} | 
  {\cal H}_0|\Psi_{\rm A}^{{\bf k}_{\rm F}+{\bf k}} \rangle = 
  \gamma_0 \frac{3a_{\rm cc}}{2} (k_x + ik_y)
  + \cdots,
 \end{split}
 \label{eq:H0_K}
\end{align}
where we used $\langle \Psi_{\rm A}^{{\bf k}_{\rm F}} | 
{\cal H}_0|\Psi_{\rm B}^{{\bf k}_{\rm F}} \rangle 
=-\gamma_0 f({\bf k}_{\rm F})=0$.
We neglect the correction indicated by 
$\cdots$ in Eq.~(\ref{eq:H0_K}) 
which is of order of ${\cal O}(k^2)$
because we only consider ${\bf k}$-states near the K point,
namely $|{\bf k}| \ll |{\bf k}_{\rm F}|$.

From Eq.~(\ref{eq:H0_K}), we see that Eq.~(\ref{eq:H_AB})
is approximated by
\begin{align}
 & E({\bf k}_{\rm F}+{\bf k})
 \begin{pmatrix}
  | \Psi_{\rm A}^{{\bf k}_{\rm F}+ {\bf k}} \rangle \cr
  | \Psi_{\rm B}^{{\bf k}_{\rm F}+ {\bf k}} \rangle
 \end{pmatrix}
 \nn \\
 &= \frac{3 \gamma_0 a_{\rm cc}}{2}
 \begin{pmatrix}
  0 & k_x - ik_y \cr
  k_x + ik_y & 0 
 \end{pmatrix}
 \begin{pmatrix}
  | \Psi_{\rm A}^{{\bf k}_{\rm F}+ {\bf k}} \rangle \cr
  | \Psi_{\rm B}^{{\bf k}_{\rm F}+ {\bf k}} \rangle
 \end{pmatrix}.
 \label{eq:H_AB_n}
\end{align}
By introducing
the Fermi velocity as $v_{\rm F} = 3 \gamma_0 a_{\rm cc}/2\hbar$,
the momentum operator $\hat{\bf p}=-i\hbar \nabla$, 
and the Pauli matrix $\bsigma=(\sigma_x,\sigma_y)$,
we obtain the effective-mass Hamiltonian 
as $v_{\rm F} \bsigma \cdot \hat{\mathbf{p}}$
which is given by Eq.~(\ref{eq:H0K}).

A lattice deformation induces 
a local modification of the nearest-neighbor hopping integral
as $-\gamma_0 \to -\gamma_0+\delta \gamma_0^a({\bf r}_i)$ 
($a=1,2,3$) (see Fig.~\ref{fig:graphene}(a)).
We define this perturbation as Eq.~(\ref{eq:H1}).
The off-site matrix element of ${\cal H}_1$ 
with respect to the Bloch wavefunction of Eq.~(\ref{eq:Bloch})
is given by
\begin{align}
 \begin{split}
  & \langle \Psi_{\rm A}^{{\bf k}+\delta {\bf k}} |{\cal H}_1|
  \Psi_{\rm B}^{{\bf k}} \rangle \\
  &= \frac{1}{N_u} \sum_{i \in {\rm A}} \sum_{a=1,2,3} 
  \delta \gamma^a_0({\bf r}_i) f_a({\bf k}) e^{-i\delta {\bf k}
  \cdot {\bf r}_i}, \\
  & \langle \Psi_{\rm B}^{{\bf k}+\delta {\bf k}} |{\cal H}_1|
  \Psi_{\rm A}^{{\bf k}} \rangle \\
  &= \frac{1}{N_u} \sum_{i \in {\rm A}} \sum_{a=1,2,3} 
  \delta \gamma^a_0({\bf r}_i) f_a({\bf k})^* 
  e^{-i\delta {\bf k}\cdot ({\bf r}_i+{\bf R}_a)}.
 \end{split}
 \label{eq:H1_bb}
\end{align}
By changing ${\bf k}$ in Eq.~(\ref{eq:H1_bb}) 
to ${\bf k}_{\rm F}+{\bf k}$ and 
using $f_a({\bf k}_{\rm F}+{\bf k}) = f_a({\bf k}_{\rm F}) +
if_a({\bf k}_{\rm F}){\bf k}\cdot {\bf R}_a + \cdots$,
we see that
\begin{align}
 \begin{split}
  & \langle \Psi_{\rm A}^{{\bf k}_{\rm F}+{\bf k}+\delta {\bf k}} |
  {\cal H}_1| \Psi_{\rm B}^{{\bf k}_{\rm F}+{\bf k}} \rangle \\
  & = \frac{1}{N_u} \sum_{i \in {\rm A}} \sum_{a=1,2,3} 
  \delta \gamma_a({\bf r}_i) f_a({\bf k}_{\rm F}) 
  e^{-i\delta {\bf k} \cdot {\bf r}_i}+\cdots, \\
  & \langle \Psi_{\rm B}^{{\bf k}_{\rm F}+{\bf k}+\delta {\bf k}} |
  {\cal H}_1| \Psi_{\rm A}^{{\bf k}_{\rm F}+{\bf k}} \rangle \\
  &= \frac{1}{N_u} \sum_{i \in {\rm A}} \sum_{a=1,2,3} 
  \delta \gamma_a({\bf r}_i) f_a({\bf k}_{\rm F})^*
  e^{-i\delta {\bf k} \cdot {\bf r}_i}+\cdots.
 \end{split}
 \label{eq:g-basic}
\end{align}
The correction indicated by $\cdots$ in Eq.~(\ref{eq:g-basic})
is negligible when $|{\bf k}| \ll |{\bf k}_{\rm F}|$ and 
$|\delta {\bf k}| \ll |{\bf k}_{\rm F}|$.
Substituting $f_1({\bf k}_{\rm F}) = 1$, 
$f_2({\bf k}_{\rm F})=e^{-i\frac{2\pi}{3}}$ and 
$f_3({\bf k}_{\rm F})=e^{+i\frac{2\pi}{3}}$ 
into Eq.(\ref{eq:g-basic}),
we get 
\begin{align}
 \begin{split}
  & \langle \Psi_{\rm A}^{{\bf k}_{\rm F}+{\bf k} + \delta {\bf k}} 
  |{\cal H}_1| \Psi_{\rm B}^{{\bf k}_{\rm F}+{\bf k}} \rangle \\
  &= \frac{v_{\rm F}}{N_{u}} \sum_{i \in {\rm A}}
  \left\{
  A_x({\bf r}_i) -i A_y({\bf r}_i) \right\}
  e^{-i\delta {\bf k} \cdot {\bf r}_i}, \\
  & \langle \Psi_{\rm B}^{{\bf k}_{\rm F}+{\bf k} + \delta {\bf k}} 
  |{\cal H}_1| \Psi_{\rm A}^{{\bf k}_{\rm F}+{\bf k}} \rangle \\
  &= \frac{v_{\rm F}}{N_{u}} \sum_{i \in {\rm A}}
  \left\{
  A_x({\bf r}_i) +i A_y({\bf r}_i) \right\}
  e^{-i\delta {\bf k} \cdot {\bf r}_i},
 \end{split}
 \label{eq:off-ele}
\end{align}
where
${\bf A}({\bf r})=(A_x({\bf r}),A_y({\bf r}))$
is defined from 
$\delta \gamma^a_0({\bf r})$ ($a=1,2,3$) as
\begin{align}
 \begin{split}
  & v_{\rm F} A_x({\bf r}) = \delta \gamma^1_0({\bf r})
  - \frac{1}{2} \left( \delta \gamma^2_0({\bf r}) +
  \delta \gamma^3_0({\bf r}) \right), \\
  & v_{\rm F} A_y({\bf r}) = \frac{\sqrt{3}}{2} 
  \left( \delta \gamma^2_0({\bf r}) -
  \delta \gamma^3_0({\bf r}) \right).
 \end{split}
\end{align}
Equation~(\ref{eq:off-ele}) shows that
${\cal H}_1$ appears as 
$v_{\rm F} \bsigma \cdot {\bf A}({\bf r})$
in the effective-mass Hamiltonian.
Thus, the total Hamiltonian becomes
$v_{\rm F} \bsigma \cdot (\hat{\bf p}+{\bf A}({\bf r}))$,
which is given by Eq.~(\ref{eq:W}).

\section{Chirality dependence of the electron-phonon
 interaction}\label{app:chi}

Next
we derive the el-ph interaction for the RBM 
in an $(n,m)$ SWNT.

\begin{figure}[htbp]
 \begin{center}
  \includegraphics[scale=0.45]{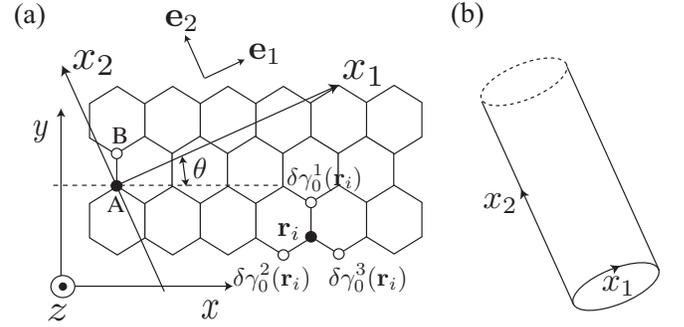}
 \end{center}
 \caption{
 (a) We define the coordinate system $(x_1,x_2)$ for graphene.
 ${\bf e}_1$ (${\bf e}_2$) is the dimensionless unit vector for the
 $x_1$-axis ($x_2$-axis). 
 (b) $x_1$ ($x_2$) is the coordinate around (along) a tube axis.
 The periodic boundary condition along the $x_1$ axis corresponds to
 a carbon nanotube with chiral angle $\theta$.
 }
 \label{fig:graphene-cnt}
\end{figure}

A straightforward method to derive the effective-mass Hamiltonian
for an $(n,m)$ nanotube is to represent a vector 
in terms of unit vectors
${\bf e}_1={\bf C}_h/|{\bf C}_h|$ and
${\bf e}_2={\bf T}/|{\bf T}|$ 
instead of ${\bf e}_x$ and ${\bf e}_y$ 
(see Fig.~\ref{fig:graphene-cnt}(a)).
Here ${\bf C}_h$ and ${\bf T}$
are the chiral and translational vectors, respectively.~\cite{saito98book}
The relationship between $({\bf e}_1,{\bf e}_2)$
and $({\bf e}_x,{\bf e}_y)$ is given by the chiral angle as
\begin{align}
 \begin{pmatrix}
  {\bf e}_x \cr {\bf e}_y
 \end{pmatrix}
 = 
 \begin{pmatrix}
  \cos \theta & - \sin \theta \cr
  \sin \theta & \cos \theta
 \end{pmatrix}
 \begin{pmatrix}
  {\bf e}_1 \cr {\bf e}_2
 \end{pmatrix}.
 \label{eq:coor_tran}
\end{align}

Using Eq.~(\ref{eq:coor_tran}), 
we represent ${\bf R}_a$ ($a=1,2,3$)
in terms of ${\bf e}_1$, ${\bf e}_2$, and $\theta$ as
\begin{align}
 \begin{split}
  & \frac{{\bf R}_1}{a_{\rm cc}}=
  \sin \theta {\bf e}_1 + \cos \theta {\bf e}_2, \\
  & \frac{{\bf R}_2}{a_{\rm cc}}=
  \left(- \frac{\sqrt{3}}{2} \cos \theta - \frac{1}{2} \sin \theta
  \right){\bf e}_1 
  + \left(\frac{\sqrt{3}}{2} \sin \theta - \frac{1}{2} \cos \theta
  \right){\bf e}_2,
  \\
  & \frac{{\bf R}_3}{a_{\rm cc}}=
  \left( \frac{\sqrt{3}}{2} \cos \theta - \frac{1}{2} \sin \theta
  \right){\bf e}_1 
  + \left(-\frac{\sqrt{3}}{2} \sin \theta - \frac{1}{2} \cos \theta
  \right){\bf e}_2.
 \end{split}
 \label{eq:Ra_new}
\end{align}
Then, 
by following the same procedure to get Eq.~(\ref{eq:H0_K})
of Appendix~\ref{app:W+A}, 
we obtain 
\begin{align}
 \begin{split}
  & \langle \Psi_{\rm A}^{{\bf k}_{\rm F}+{\bf k}} | 
  {\cal H}_0|\Psi_{\rm B}^{{\bf k}_{\rm F}+{\bf k}} \rangle = 
  e^{-i\theta} \gamma_0 \frac{3a_{\rm cc}}{2} (k_1 - ik_2)
  + \cdots, \\
  & \langle \Psi_{\rm B}^{{\bf k}_{\rm F}+{\bf k}} | 
  {\cal H}_0|\Psi_{\rm A}^{{\bf k}_{\rm F}+{\bf k}} \rangle = 
  e^{+i\theta} \gamma_0 \frac{3a_{\rm cc}}{2} (k_1 + ik_2)
  + \cdots,
 \end{split}
 \label{eq:H0_K_c}
\end{align}
where we denote the wave vector in the direction 
around (along) the tube axis $k_1$ ($k_2$) that is
${\bf k}=k_1 {\bf e}_1 + k_2 {\bf e}_2$
(see Fig.~\ref{fig:graphene-cnt}(a) and (b)).

Let us introduce a new Bloch wave function 
which is defined by adding a chiral angle dependent phase 
to the original Bloch wave function of Eq.~(\ref{eq:Bloch}) as
\begin{align}
 \begin{split}
  & |\Psi_{\rm A}^{\bf k}(\theta) \rangle= 
  e^{-i\theta/2}|\Psi_{\rm A}^{\bf k} \rangle, \\
  & |\Psi_{\rm B}^{\bf k}(\theta) \rangle= 
  e^{+i\theta/2}|\Psi_{\rm B}^{\bf k} \rangle.
 \end{split}
 \label{eq:bloch_new}
\end{align}
Then the effective-mass Hamiltonian becomes
\begin{align}
 v_{\rm F} (\sigma_x p_1 + \sigma_y p_2),
\end{align}
and Eq.~(\ref{eq:off-ele}) becomes
\begin{align}
 \begin{split}
  & \langle \Psi_{\rm A}^{{\bf k}_{\rm F}+{\bf k} + \delta {\bf k}}(\theta)
  |{\cal H}_1| \Psi_{\rm B}^{{\bf k}_{\rm F}+{\bf k}}(\theta) \rangle \\
  &= \frac{v_{\rm F}}{N_{u}} \sum_{i \in {\rm A}} e^{+i\theta}
  \left\{
  A_x({\bf r}_i) -i A_y({\bf r}_i) \right\}
  e^{-i\delta {\bf k} \cdot {\bf r}_i}, \\
  & \langle \Psi_{\rm B}^{{\bf k}_{\rm F}+{\bf k} + \delta {\bf k}} (\theta)
  |{\cal H}_1| \Psi_{\rm A}^{{\bf k}_{\rm F}+{\bf k}}(\theta) \rangle \\
  &= \frac{v_{\rm F}}{N_{u}} \sum_{i \in {\rm A}} e^{-i\theta}
  \left\{
  A_x({\bf r}_i) +i A_y({\bf r}_i) \right\}
  e^{-i\delta {\bf k} \cdot {\bf r}_i}.
 \end{split}
\end{align}
Therefore,
by introducing $A_1({\bf r})$ and $A_2({\bf r})$ 
which are defined by
\begin{align}
 \begin{pmatrix}
  A_1({\bf r}) \cr A_2({\bf r})
 \end{pmatrix}
 = 
 \begin{pmatrix}
  \cos \theta & \sin \theta \cr
  -\sin \theta & \cos \theta
 \end{pmatrix}
 \begin{pmatrix}
  A_x({\bf r}) \cr A_y({\bf r})
 \end{pmatrix},
 \label{eq:A_general}
\end{align}
we see that the off-site interaction can be written as
\begin{align}
 {\cal H}_1^{\rm K} = \sigma_x A_1({\bf r}) + \sigma_y A_2({\bf r}).
\end{align}
The Hamiltonian for a chiral SWNT can be represented by 
\begin{align}
 v_{\rm F} \bsigma \cdot ({\bf p}+{\bf A}({\bf r}))
\end{align}
where ${\bf p}=(p_1,p_2)$ and 
${\bf A}({\bf r})=(A_1({\bf r}),A_2({\bf r}))$.

To determine $A_1({\bf r})$ and $A_2({\bf r})$,
it is necessary to represent 
$A_x({\bf r})$ and $A_y({\bf r})$
in the new coordinate system.
We shall do this for obtaining the el-ph interaction for the RBM.
Putting Eq.~(\ref{eq:Ra_new}) into 
\begin{align}
 \delta \gamma^a_0({\bf r}) 
 \approx \frac{g_{\rm off}}{\ell a_{\rm cc}}
 {\bf R}_a \cdot 
 \left( ({\bf R}_a \cdot \nabla) {\bf s}({\bf r}) \right)
\end{align}
of Eq.~(\ref{eq:dg_ac}),
we get using Eq.~(\ref{eq:A})
\begin{widetext}
\begin{align}
 \begin{split}
  & v_{\rm F}A_x({\bf r}) = \frac{g_{\rm off}}{2} 
  \left\{
  \cos2\theta \left(
  -\frac{\partial s_1({\bf r})}{\partial x_1}+
  \frac{\partial s_{2}({\bf r})}{\partial x_2} \right)
  + \sin 2\theta \left(
  \frac{\partial s_1({\bf r})}{\partial x_2}+
  \frac{\partial s_2({\bf r})}{\partial x_1} \right)
  \right\}, \\
  & v_{\rm F}A_y({\bf r}) = \frac{g_{\rm off}}{2}  
  \left\{
  -\sin 2\theta \left(
  -  \frac{\partial s_1({\bf r})}{\partial x_1}+
  \frac{\partial s_{2}({\bf r})}{\partial x_2} \right)
  + \cos 2\theta \left(
  \frac{\partial s_1({\bf r})}{\partial x_2}+
  \frac{\partial s_2({\bf r})}{\partial x_1} \right)
  \right\}.
 \end{split}
 \label{eq:A_new}
\end{align}
\end{widetext}
Inserting Eq.~(\ref{eq:A_new}) into Eq.~(\ref{eq:A_general}), 
we see that 
\begin{widetext}
\begin{align}
 v_{\rm F}
 \begin{pmatrix}
  A_1({\bf r}) \cr A_2({\bf r})
 \end{pmatrix}
 =
 \frac{g_{\rm off}}{2}
 \begin{pmatrix}
  \cos 3\theta & \sin 3\theta \cr
  -\sin 3\theta & \cos 3\theta
 \end{pmatrix}
 \begin{pmatrix}
  \displaystyle
  -\frac{\partial s_1({\bf r})}{\partial x_1}+
  \frac{\partial s_2({\bf r})}{\partial x_2} \cr
  \displaystyle
  \frac{\partial s_1({\bf r})}{\partial x_2}+
  \frac{\partial s_2({\bf r})}{\partial x_1}
 \end{pmatrix}.
    \label{eq:A_chiral}
\end{align}
\end{widetext}
The off-site el-ph interaction for the RBM of a chiral SWNT
is given by replacing 
\begin{align}
 \frac{\partial s_1({\bf r})}{\partial x_1} \to
 \frac{\partial s_1({\bf r})}{\partial x_1} + 
 \frac{s_z}{r}
\end{align}
in Eq.~(\ref{eq:A_chiral}) as 
we have explained in Sec.~\ref{sec:hami}.
For the $\Gamma$ point RBM, since 
$\partial s_1({\bf r})/\partial x_i=0$ and
$\partial s_2({\bf r})/\partial x_i=0$ ($i=1,2$),
we obtain
$v_{\rm F}A_1=-g_{\rm off}(s_z/d_t)\cos 3\theta$ and
$v_{\rm F}A_2=g_{\rm off}(s_z/d_t)\sin 3\theta$.
This gives the el-ph interaction written in Eq.~(\ref{eq:rbm_int}).




\end{document}